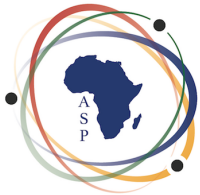

# African School of Fundamental Physics and Applications

### ASP Activity Report, 2019-2021

Kétévi A. Assamagan [1] (BNL), Bobby Acharya (ICTP),

Temitope Adenuga (University of Ibadan, Nigeria),

Mohamed Chabab (Cadi Ayyad University, Morocco),

Kenneth Cecire (University of Notre Dame),

Simon H. Connell (University of Johannesburg), Anne E. Dabrowski (CERN),

Christine Darve (ESS), Farida Fassi (Mohammed V University, Morocco),

Jonathan R. Ellis (University of London), Fernando Ferroni (INFN-GSSI),

Mounia Laassiri (Mohammed V University, Morocco), Steve G. Muanza (CNRS-IN2P3)

## Abstract

The sixth edition of the African School of Fundamental and Applied Physics (ASP) was planned for Morocco in July 2020 and was referred to as ASP2020. Preparations were at an advanced stage when ASP2020 was postponed because of the COVID-19 pandemic. The three-week event was restructured into two activities in 2021 — an online event on July 19-30, 2021 and a hybrid event on December 12-18, 2021 — and was renamed ASP2021. At the beginning of the COVID-19 pandemic, an online lecture series was integrated into the ASP activities. The ASP mentorship program, which consists of online engagements between lecturers and assigned mentees, continued in this way. ASP alumni studied one year of COVID-19 data of ten African countries to offer insights into pandemic containment measures. In this note, we report on ASP activities since the last in-person edition of ASP in 2018 in Namibia.

## 1) Introduction

International cooperation forms the common denominator of today's culture of scientific activities. However, in many scientific disciplines and especially in fundamental and applied physics the cooperation among African countries and between them and the rest of the world is not well developed. This is especially the case for sub-Saharan Africa, which is one of the most rapidly developing regions in the world with great educational needs. In order to extend the existing international scientific ties to this geographical zone, we have established a biennial African School of Physics [1] with a focus on fundamental and applied physics.

The ASP series started in 2010 in South Africa, then continued to Ghana (2012), Senegal (2014), Rwanda (2016), and Namibia (2018) [2-6]. The 2020 edition of ASP was planned to be in Morocco; however, because of the COVID-19 pandemic, it was organized online in July 2021. The ASP is based on the close interplay between theoretical, experimental, and applied physics, as well as computing. It covers a wide range of topics in fundamental and applied physics. About

---

[1] Corresponding author Email: ketevi@bnl.gov

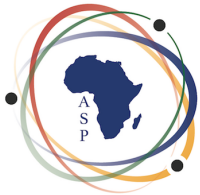 # African School of Fundamental Physics and Applications

eighty students are selected from all over Africa, from upwards of four hundred applications in each edition. International scientists are invited to prepare and deliver lectures according to the proposed topics considering the diverse levels and backgrounds of the students. The duration of the school allows for extensive networking between participants. A one-week training workshop for about seventy high school teachers and a one-week outreach for over fifteen hundred high school pupils are included in the program. These workshops introduce teachers and pupils to the excitement of contemporary physics in accessible and replicable ways.

Research institutions, universities, government agencies, and foundations have sponsored ASP. The success of the school is sufficiently encouraging to provide motivation for a review of the ASP goals and for consideration of mechanisms that would make it sustainable. The central long-term objective of ASP is to help improve physics education in Africa and in doing so, to contribute in a significant way to the development of science and technology on this continent. We believe that maintaining the leadership of the organization of ASP in partnership with other interested institutes and African governments and policy makers presents a unique opportunity to pioneer the scientific and technological development of a region of more than one billion people with large unmet needs but vast human potential.  What is needed at this time to ensure the future of ASP and the success of its mission are partners that can provide sustained support for the participation of African students, teachers and pupils.

The continuous support for the participation of African students, pupils, teachers and research faculties in future ASP can be realized in various ways: direct financial support to the budget of the school to cover participant travels; travel support for ASP organizers / lecturers in the activities that enhance the reach and coverage of the ASP; or travel coverage for ASP alumni to spend 3-6 months at international research labs.

The objective of ASP is achieved through an outreach effort, an increased awareness of the potential of high-quality training offered by large scale experiments in the context of various scientific disciplines, and a system of networking on the international scale. There is a strong alignment between the mission and the vision of African governments and policy makers on education and capacity building and their programs with the goals of the ASP. The ASP is committed to include African governments in the planning, in order to take advantage of aspects such as consolidating agreements and their goals, building on synergy with other programs, improving the sustainability and impact of capacity development and improving the measurement and visibility of the impact. By working with African governments and policy makers on education, ASP seeks to promote a culture of science that creates an attractive environment for African students, thus encouraging their retention within Africa. ASP promotes sustainable scientific development in Africa by building a network between African and international researchers for increased collaborative research and shared expertise—see Appendix A for details.

The success of ASP is due to the sponsorship of supporting institutes, the dedication of the organizing committee, the lecturers, and the students. Many students in Africa face challenges in

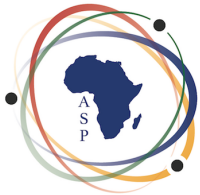

# African School of Fundamental Physics and Applications

terms of the logistical support, the quality of education and the opportunity for higher education. It is often the case in Africa that even the best students do not have the needed support to succeed or to acquire the necessary skills to be competitive at an international level. It is important to help resolve some of the challenges to improve physics education and research in Africa. ASP serves at least two purposes: it provides a template for solving educational challenges, and offers opportunities for networking, which helps prepare the students to find practical answers to many issues.

In this report, we present ASP activities since the last edition in Namibia in 2018 and the changes made as a result of the COVID-19 pandemic.

## 2) Short-term visits for research

At the conclusion of ASP2018 in Namibia, we sought funding to complement the ASP program with short-term visits to international research laboratories for research. In this program, depending on the availability of funding, a number of outstanding ASP alumni are selected to spend 3-6 months at international research laboratories that sponsor ASP. Nine ASP alumni went to Brookhaven National Laboratory (BNL) between July and December 2019. They were placed within different research groups and assigned research advisors according to their majors and physics interests. This program was supported by funds from DOE, the BNL Diversity Office, Nuclear & Particle Physics Diversity Council and the departments and research groups that hosted participants. At ASP2016, Christelle Ekosso won the student poster contest, which entitled her to 1000 Euros [5]; that award money was used to cover Christelle Ekosso travel to BNL; Hassnae El Jarrari was at Jefferson Lab for the Hampton University Graduate Students program and had a returned air ticket to Morocco. All other program expenses were covered by DOE and BNL, including Hassnae El Jarrari's travel to BNL and the change fee of her air ticket return date. Figure 1 shows some of the ASP alumni that participated in this program.

Somiéalo Azote and Mounia Laassiri received their PhD degree before they went to BNL. After this short-term visit, Diallo Boye, Christelle Ekosso and Heba Sami Abdulraham completed their PhD and are now post-doc; Raymond Yogo, Jesutofunmi Fajemisin and Yes Kini are pursuing graduate studies. During their visits, Yves Kini, worked under the mentorship of Dr. Peter Denton and Dr. Mary Bishai of BNL on neutrino physics. Their collaboration continued beyond the short-term visit and led to a published paper [7]. In August 2021, Yves Kini was awarded the inaugural Augustus Prince Scholar Award for this work. During these visits, Mounia Laassiri was invited to give a talk on ASP at the American Physical Society Division of Particles and Fields (DPF) annual meeting in Boston; the contribution to the DFP proceedings is shown in Ref. [8].

We expect this program to continue biennially during consecutive term schools and to extend to other research laboratories. Because of the COVID-19 pandemic, no visits were arranged in 2021; the program will resume when the pandemic is under control and travels and visits are permitted.

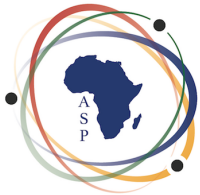

# African School of Fundamental Physics and Applications

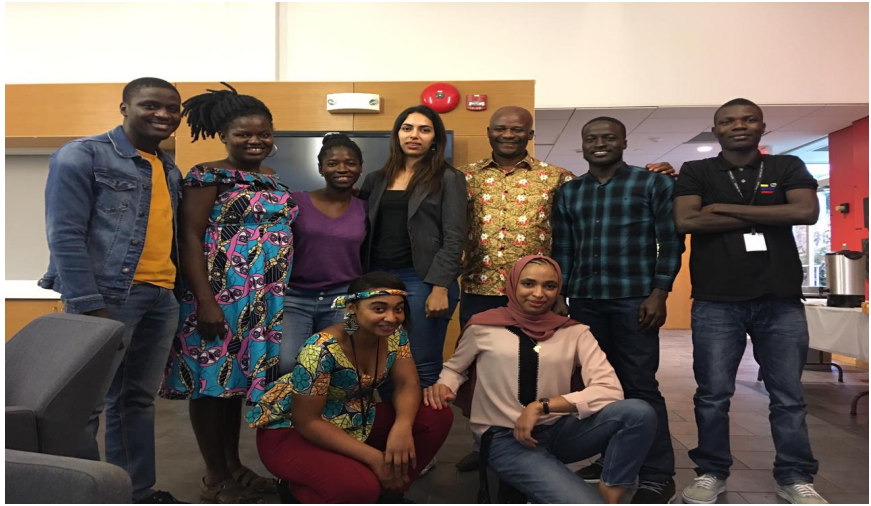

**Figure 1:** From left to right: in front, Dr. Christelle Ekosso (Cameroon), Dr. Mounia Laassiri (Morocco); standing, Dr. Diallo Boye (Senegal), Dr. Somiealo Azote (Togo), Jesutofunmi Fajemisin (Nigeria), Hassnae El Jarrari (Morocco), Dr. Kétévi A. Assamagan, Raymond Yogo (Kenya), and Yves Kini (Burkina Faso). Heba Sami Abdulrahman (Egypt), not in the figure, arrived in September 2019.

The short-term visits are an important addition to the ASP program to offer ASP alumni international exposure to work with mentors and leading experts in their fields, to gain valuable research experience and to be well prepared for graduate or post-doctoral studies. Feedback on this program is compiled in Ref. [9].

## 3) ASP2020

In December 2017, we selected Morocco to host the sixth edition of the ASP at the Cadi Ayyad University in Marrakesh in July 2020. In April 2019, the International Organizing Committee (IOC) made a site visit to Morocco, and met with members of the Local Organizing Committee (LOC), university and government officials in Rabat and Marrakesh. The objective of the ASP site visit is to discuss the logistics of the upcoming school and set the stage for concerted preparation toward the event.

We opened student applications for a period of three months starting October 15, 2019. We received upwards of 450 applications by January 15, 2020. A selection committee of 28 members from the IOC, the LOC and the international lecturers reviewed the applications and by February 15, 2020, selected 70 applicants with 31 additional good candidates on the waiting list. By that time, we had to make a 30% down payment to secure student lodging and catering services. Towards the end of February 2020, as we were setting to start travel arrangements for students, the COVID-19 pandemic took an unprecedented worldwide dimension and ASP2020 was postponed; we decided to come back to Morocco for ASP2024.

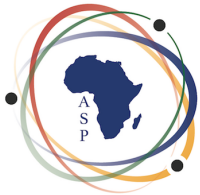

# African School of Fundamental Physics and Applications

**4) Online seminars**

From May 2020, we extended ASP activities with the addition of weekly online seminars, open to not only ASP alumni, but also interested folks worldwide. We invite former ASP lecturers and international research faculties in a wide range of topics as shown in Ref. [10]. We encourage ASP alumni that are at the level of graduate studies or above to give seminars on their research work. During the period of December 2020 to June 2021, we counted 12 ASP alumni that completed their PhD degree.

**5) Mentorship program**

Mentorship is a continuous activity even when there is no term school. Every two years, in-between consecutive biennial schools, we select ASP alumni that are at the level of independent research. We solicit volunteer mentors from ASP lecturers and organizers. We pair up the mentees with the mentors considering common research interests. Since we rely on available lecturers to serve as mentors and the students come with diverse physics backgrounds, we are not always able to do the pairings in the same physics backgrounds. Nevertheless, the accessibility of lecturers to mentor ASP students is a valuable addition to the ASP program.

In October 2020, we made a call for the second around of ASP structured mentorship and we selected and paired up 20 mentees in February 2021. The current mentorship cycle will run until 2023; then, we will select new mentees and seek new mentors. Halfway through the current cycle and before the new cycle begins, we will solicit feedback from the mentors and mentees to improve the program.

**6) COVID-19 Analysis**

In April 2020, we suggested to ASP alumni to study COVID-19 data of their own country to understand the time evolution of the basic reproduction, $R_0$. The understanding of the time-dependent $R_0$ should help with containment measures. A number of alumni volunteered and we formed a study group that met weekly. The analysis team members are shown in Figure 2. We studied one year of COVID-19 data of ten African countries: Benin, Cameroon, Ghana, Kenya, Madagascar, Mozambique, Rwanda, South Africa, Togo and Zambia. The countries considered are the countries of residence or citizenship of the volunteers. These analyses resulted in two published papers [11].

**7) Activities in 2021**

We continued the weekly online seminars until June 2021; we expect to resume these seminars in September 2021. We also continued the COVID-19 data analysis until April 2021; by that time, we had analyzed one year on COVID-19 data of the ten African countries mentioned in Section 6.

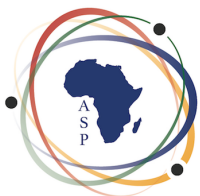

# African School of Fundamental Physics and Applications

In 2020, the term ASP, which is a three-week in-person event, was cancelled. In 2021, ASP2020 was reorganized as two distinct events: a two-week fully virtual event on July 19-30, 2021 and a hybrid event on December 12-18, 2021.

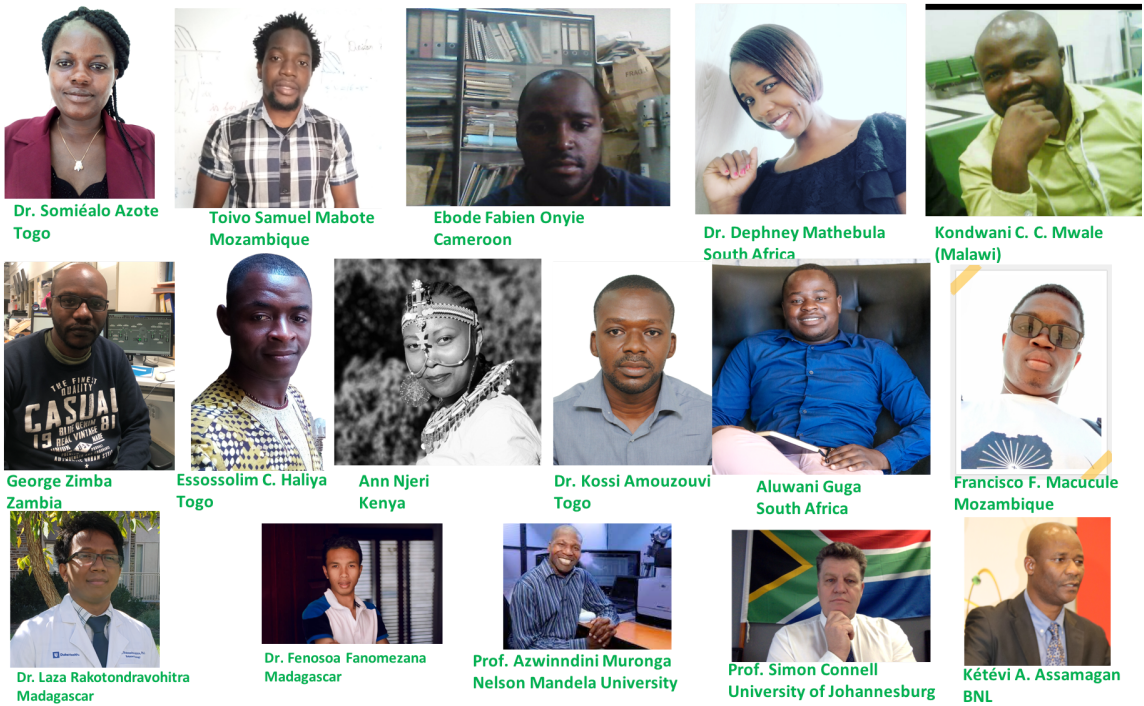

**Figure 2:** ASP COVID-19 analysis team, made of volunteer ASP alumni, lecturers and organizers who responded to the call to study the COVID-19 pandemic evolution in Africa.

## 7.1) Virtual ASP2021

The 70 students that were selected for ASP2020, as mentioned in Section 3, were eligible to attend the virtual ASP2021. Since the event was fully virtual and there was no financial burden due to travels, we extended participation to the 31 selected students on the waiting list. Furthermore, we set up a registration process for other interested folks. We received upwards of 200 registrations. During the online sessions, we noted up to 70 daily connections with an average of 56.

The online program was arranged in plenary sessions during the first week, July 19-23. These plenary sessions were attended by all the participants regardless of their academic majors. In the second week, July 26-30, we arranged two daily parallel sessions where participants were free to choose the sessions of interest. Selected posters from students were also discussed in the second week. The details of the scientific program and the lectures given during the online event—with materials presented and the recordings of the sessions—are accessible in Ref. [12]. There were a few options during the breaks:

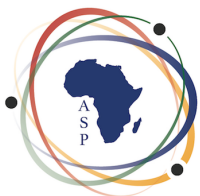

# African School of Fundamental Physics and Applications

- o   Participants were free the take the breaks;
- o   We setup an online networking forum to encourage interactions among participants;
- o   We showed ten episodes (one episode a day) of the "Science in the City" miniseries [9].

Figures 3 and 4 show respectively some of the participants (lecturers, organizers and students) and online connections during the virtual event.

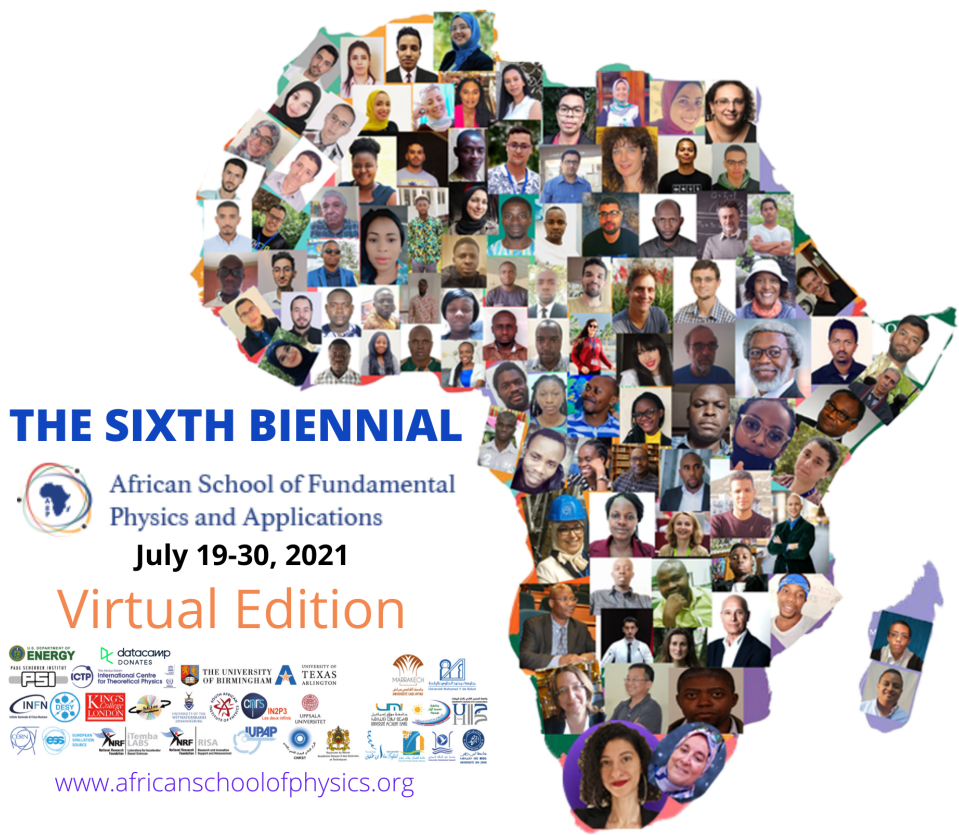

**Figure 3:** Participants in the online ASP2021.

To improve participation in the online event, we boosted internet bandwidths for selected participants. We did this by buying and uploading internet data or airtime directly to the selected students' cellular telephones through an international internet service provider. Forty ASP alumni received coverage for internet bandwidths—these were the alumni that registered for the event, provided valid telephone numbers and pledged full participation. Most of the alumni that received coverage for the internet were of ASP2020 and the rest were alumni of 2018, 2016 or 2014.

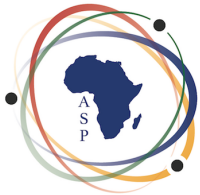

# African School of Fundamental Physics and Applications

**7.2) Hybrid ACP2021**

The second aspect of ASP2020 that was cancelled because of COVID-19 and rescheduled for 2021 is the second African Conference on Fundamental and Applied Physics (ACP2021). It is a conference that is normally integrated within the 3-week program of ASP to encourage participation of African research faculties. Since the selected students are presented at the school, they too participate in this conference. As a result, ACP is beneficial to students by offering scientific engagements and networking with international participants that are present for the conference. However, considering "online fatigue" that may result from weeks-long events, we decided to do a two-week event in July 2021 as discussed in Section 7.1, and a one-week ACP2021 on December 12-18. ACP2021 has been planned in a hybrid mode: the participants that can travel will meet at the Cadi Ayyad University in Marrakesh, Morocco; also, remote participation will be arranged. Preparation for ACP2021 is documented in Ref. [13].

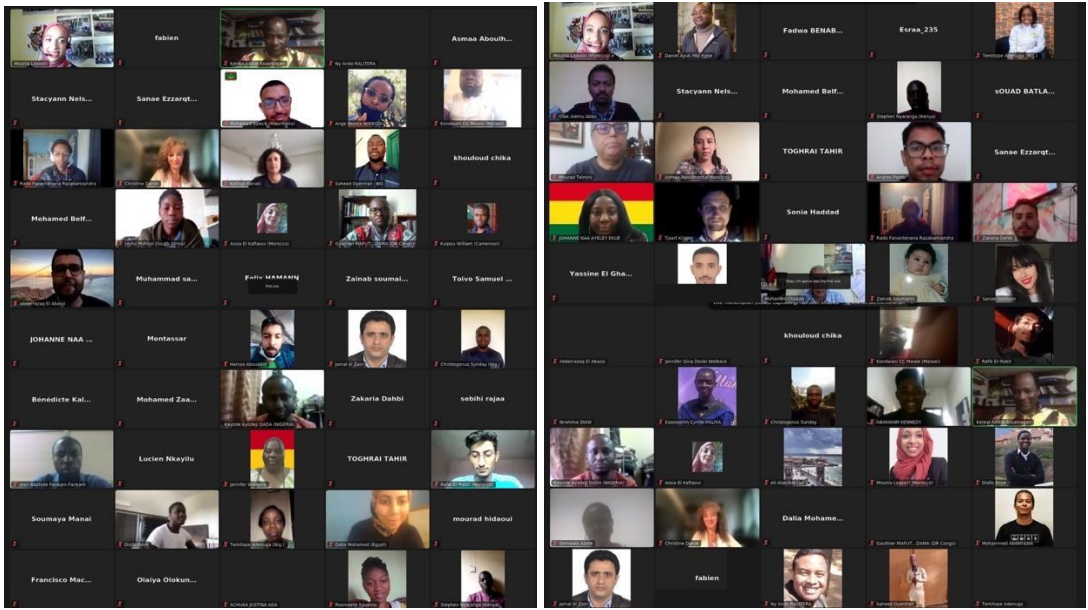

**Figure 4:** Screenshots during the online connections on July 30, 2021.

**7.3) DataCamp data analysis and coding education**

The online education platform for data analysis, DataCamp [14], donated one hundred licenses to ASP alumni for one year, starting in March 2021. All the licenses are used to gain valuable experience in data analysis. Student feedback is reported in Ref. [12].

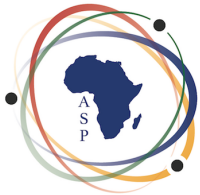

# African School of Fundamental Physics and Applications

**7.4) Survey and feedback**

Towards the end of the online event in July 2021, we collected feedback from participants in a form of a survey. Ninety-four participants offered feedback. The results are in Figures 5-8.

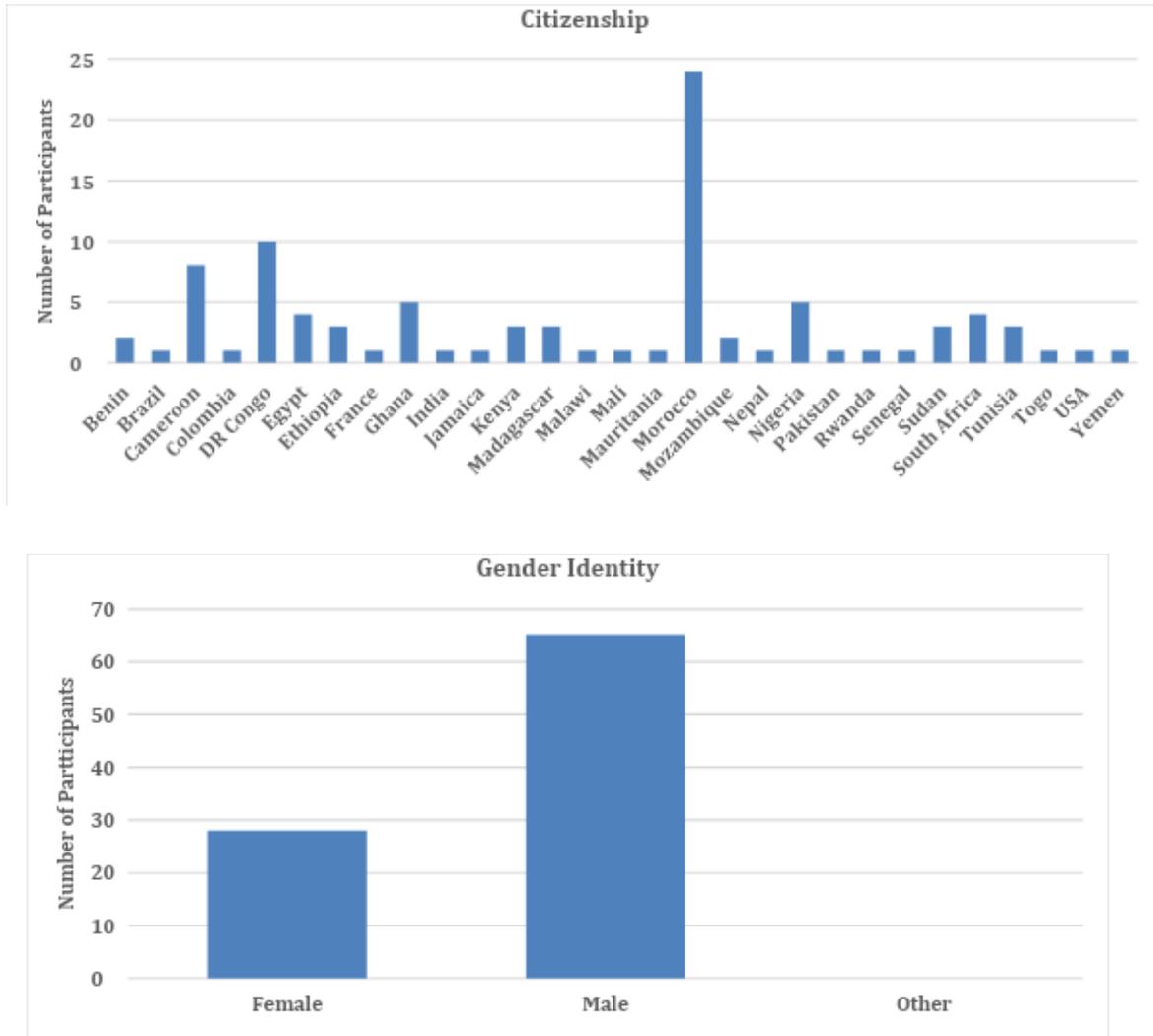

**Figure 5:** The top plot shows the countries of citizenship of the respondents. The bottom plot shows their gender identities.

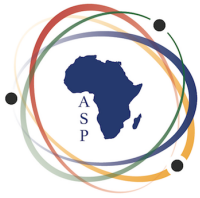

# African School of Fundamental Physics and Applications

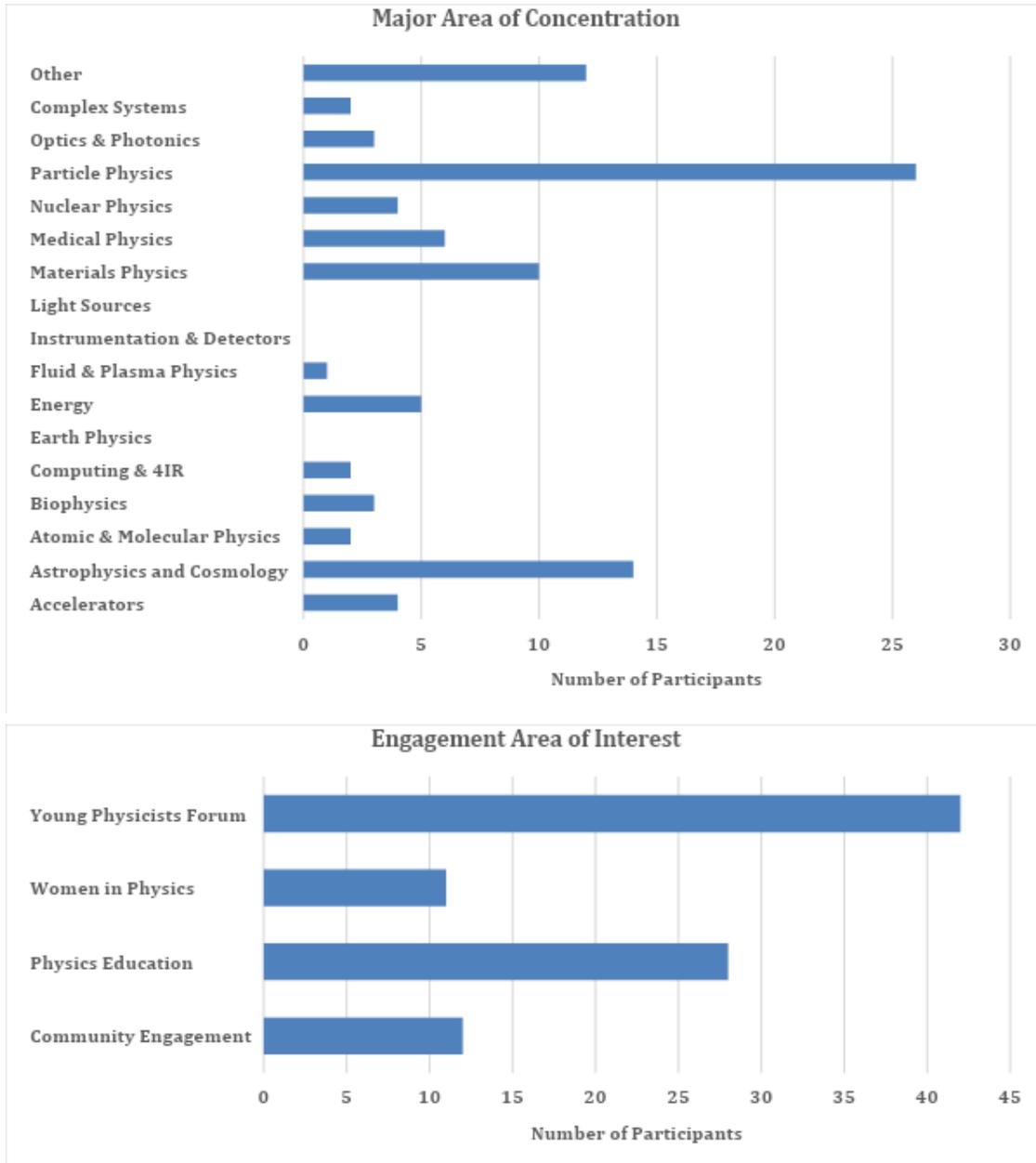

**Figure 6:** In the top panel, we show the areas of education or research concentrations of the participants; the bottom panel shows their areas of community engagements.

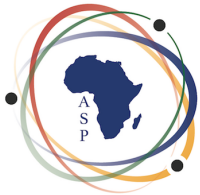

# African School of Fundamental Physics and Applications

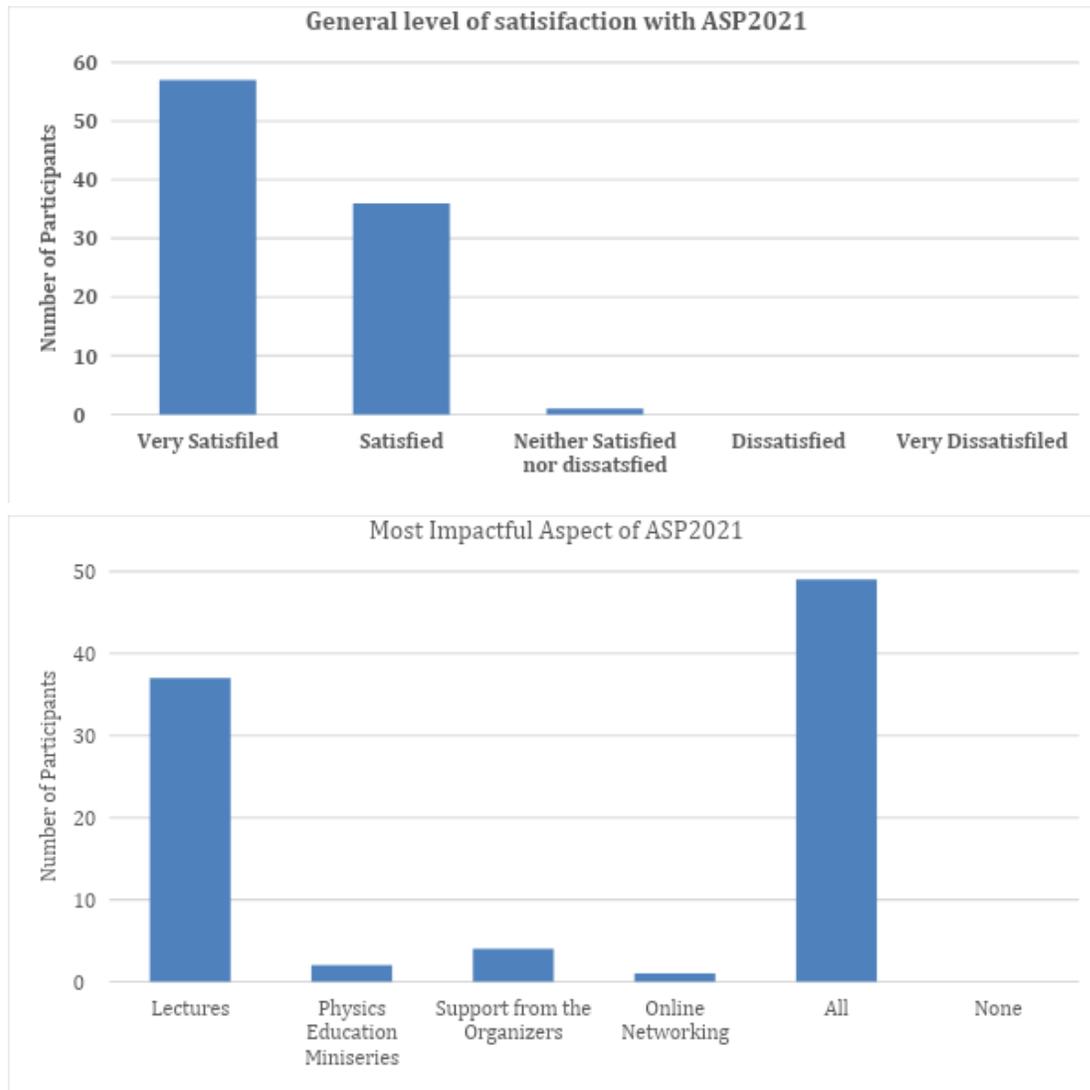

**Figure 7:** In the top plot, we asked for the level of satisfaction with the online event. In the bottom plot, participants mentioned which aspects were most impactful to them.

## 7.5) Supports and coverages

Table 1 shows the sponsors and coverages for the events described in Sections 7.1–7.3.

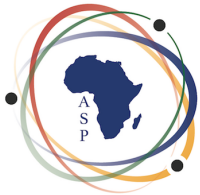

# African School of Fundamental Physics and Applications

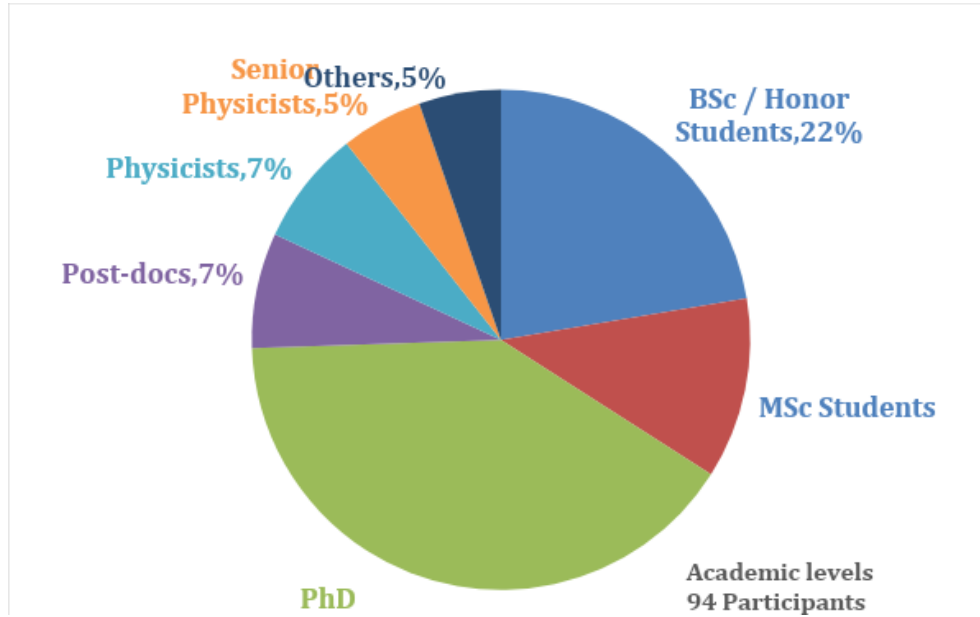

**Figure 8:** The professional standing of the respondents.

| Source | Coverage | Comment |
|---|---|---|
| DOE | Internet bandwidth for students, online networking, and online physics education miniseries | ASP2021 |
| BNL | In-person participation and travel coverages for selected ASP alumni | ACP2021 |
| CERN | Event management via Indico; Zoom online conferencing; Email lists management | ASP and ACP 2021 |
| ICTP | Payments for lecturers, early careers and student presentations; student applications management | ASP2021 |
| SAIP | Fund management; arrangements for internet bandwidths, online miniseries and online networking | ASP and ACP 2021 |
| DataCamp | 100 free licenses for online data analysis education | Free for one year |
| Variant | Online or in-person participation | ACP2021 |
| DESY | Online or in-person participation | ACP2021 |
| PSI | Online or In-person participation | ACP2021 |
| Lee Foundation | Earmarked towards events in 2022 | ASP and ACP 2022 |
| Dr. R. Lourie | Earmarked towards events in 2022 | ASP and ACP 2022 |
| Dr. J. Mitchell | Earmarked towards events in 2022 | ASP and ACP 2022 |
| Aga Khan | Earmarked towards events in 2022 | ASP and ACP 2022 |

**Table 1:** Institutional and private support for ASP2021 activities.

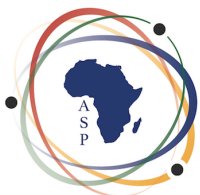

# African School of Fundamental Physics and Applications

**7.5) Online workshop for high school teachers**

In parallel to the activities mentioned in Section 7.1, we arranged an online workshop for high school teachers on July 21-23, 2021. The scientific program can be found in Ref. [15]. Figure 9 shows the geographical distribution of the registrants. The actual daily attendances were lower than the number of registrations because of internet connectivity problems.

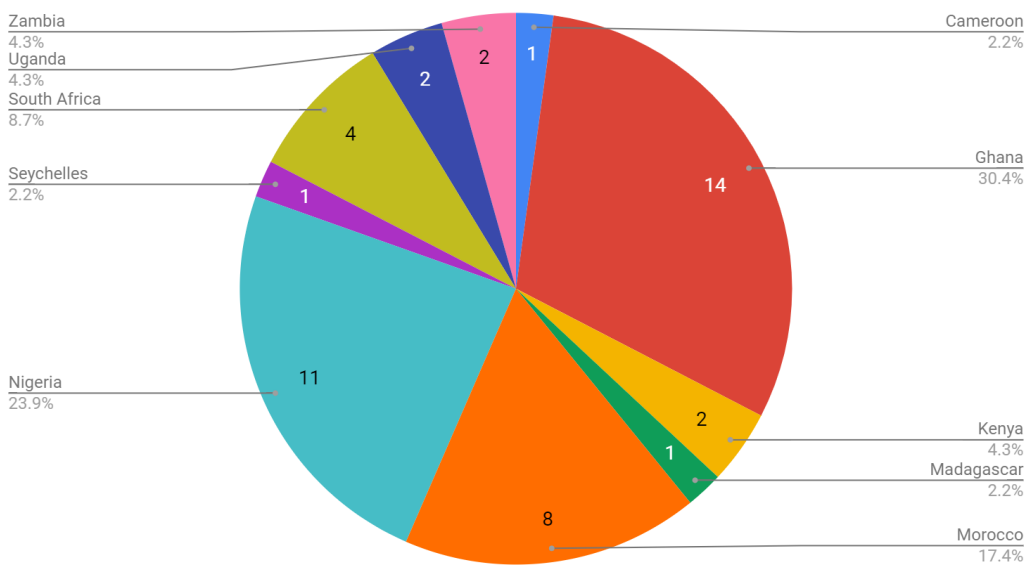

**Figure 9:** Geographical distribution of the registered participants for the online teachers' workshop.

## 8)  The ASP website

We solicited professional assistance to reorganize and improve the look-and-feel of the ASP website [1] in order to offer faster and more intuitive access to information.

**9) Partnership with the International Association of Physics Students (IAPS)**

We developed a collaboration with IAPS [16] to promote participation of ASP alumni in worldwide student activities and networking, organized and managed by students. It is expected that such a partnership will help promote IAPS in Africa and lead to a strong African student association.

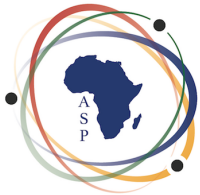

# African School of Fundamental Physics and Applications

**10) Summary**

The sixth edition of the African School of Fundamental and Applied Physics, ASP2020, was planned for July 2020 in Morocco. Preparation was at an advanced stage when the event was canceled in March 2020 because of the COVID-19 pandemic. ASP2020 was rescheduled as two events in 2021:

- o   An online series of lecturers, networking, and physics educational miniseries on July 19-30, 2021—ASP2021;
- o   A workshop for high school teachers on July 21-23, 2021;
- o   A hybrid conference planned for December 12-18, 2021 at Cadi Ayyad University in Marrakesh, Morocco—ACP2021.

ASP2021 was open to the selected students of ASP2020 and as well to other interested participants. There were over 200 registrations with up to seventy daily online connections. Feedback compiled through a survey shows that the organization and scientific program of ASP2021 were valuable to the participants. ACP2021 is an international conference on fundamental and applied physics; it is open to ASP alumni and other registered participants. It is arranged in a hybrid mode with in-person or remote participation. In addition to these activities, since ASP2018:

- o   Nine ASP alumni spent 3-6 months at Brookhaven National Laboratory for short-term visits for research in June-December 2019;
- o   We set up a series on weekly seminars since May 2020 and many of these seminars were given by ASP alumni—at the PhD level or above—on their research work;
- o   We arranged a new round of ASP structured mentorship program from March 2021;
- o   ASP alumni studied one year of COVID-19 data of ten African countries. This work resulted in two papers submitted for publication;
- o   We received one hundred free licenses from DataCamp; this allowed ASP alumni to gain valuable experience in data analysis and coding.

During the COVID-19 pandemic, ASP has continued to offer a rich variety of programs to support African students.

**11) ASP2022**

In December 2019, we selected South Africa to host the seventh edition of ASP, ASP2022,  at Nelson Mandela University. ASP2022 will be organized as two events; on July 4-8, 2022, ACP2022 will be held jointly with the South African Institute of Physics annual meeting. Then later in 2022, we will have a two-week event where the format (virtual, in-person or hybrid) will depend on the evolution of the COVID-19 pandemic.

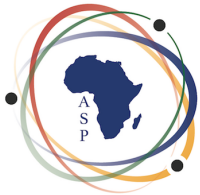

# African School of Fundamental Physics and Applications

**Acknowledgements**

We thank all the sponsors of ASP, whose support has ensured the continuity of activities even during the difficult times of the COVID-19 pandemic. Also, we thank Dr. David Bertsche for arranging the free DataCamp licenses and managing the ASP subscriptions and exercises. We appreciate the commitment of ASP lecturers and mentors.

**Appendix A**

The objective of ASP is achieved by:

- Promoting increased awareness of the potential of high quality training offered by large scale experiments in the field of fundamental physics in various scientific disciplines, and a system of networking on the international scale;
- Cultivating an appreciation of fundamental physics amongst high school teachers and learners through educational outreach programmes, mentoring and coaching;
- Hosting conferences and educational events on fundamental physics aimed at serving as a platform to engage researchers, teachers, policy makers, and industry stakeholders;
- Involving Student Alumni, ASP Trainers and African Government stakeholders in an ever-growing network whereby the ASP extends its presence beyond several active weeks per year of the actual premier School event to form a community promoting the ongoing careers of the Student Alumni, thereby furthering all the ASP aims;
- Implementing mechanisms to make fundamental physics sustainable on the African continent;
- Promoting capacity development through forging long-term partnerships with African governments, policy makers and other stakeholders;
- Promoting a culture of science that creates an attractive environment for African student alumni, thus encouraging their retention within Africa;
- Promoting sustainable scientific development in Africa by building a network between African and international researchers for increased collaborative research and shared expertise;
- Developing a new culture within African Governments of funding Science in Africa, in particular creating positions for ASP Alumni, so that high level trained human capacity developed by ASP and others is not lost to the diaspora. It is now time for Africa to lead and drive the development and decolonisation of Africa;
- Facilitating African researchers and research groups access to, and acquisition of, heavy research equipment and facilities.

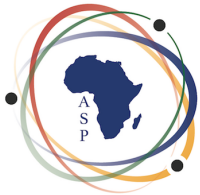

# African School of Fundamental Physics and Applications

**References**

[1] The African School of Fundamental Physics and Applications, African School of Fundamental Physics and Applications: Home

[2] Activity Report of the First Edition of ASP, 2010, http://africanschoolofphysics.web.cern.ch/2010/asp2010.pdf

[3] Activity Report of the Second Edition of ASP, 2012, https://africanschoolofphysics.web.cern.ch/asp2012/asp2012_final.pdf

[4] Activity Report of the Third Edition of ASP, 2014, https://www.africanschoolofphysics.org/wp-content/uploads/2014/11/asp2014.pdf

[5] Activity Report of the Fourth Edition of ASP, 2016, https://www.africanschoolofphysics.org/wp-content/uploads/2019/08/ASP2016-FinalReport.pdf

[6] Activity Report of the Fifth Edition of ASP, 2018, https://www.africanschoolofphysics.org/wp-content/uploads/2019/08/ASP2018.pdf

[7] Phys. Rev. D 102, 123019 (2020), https://arxiv.org/abs/2007.10334

[8] Contribution to DFP 2019 proceedings, https://arxiv.org/abs/1909.06309

[9] Building up the African physics community, https://www.symmetrymagazine.org/article/building-up-the-african-physics-community https://www.bnl.gov/newsroom/news.php?a=216732.

[10] ASP online seminars, https://www.africanschoolofphysics.org/online-lecture-series/

[11] ASP COVID-19 analyses, International Journal of Public Health and Epidemiology, Vol. 10 (8), pp. 001-016, August, 2021; https://doi.org/10.46882/IJPHE/1235; arXiv:2007.10927. Scientific African (2021), https://doi.org/10.1016/j.sciaf.2021.e00987; arXiv2104.09675.

[12] Scientific Program of the online ASP2021 on July 19-30, https://indico.cern.ch/event/812393/timetable/?view=standard

[13] The second African conference on fundamental and applied physics, ACP2021, https://indico.cern.ch/event/1060503/

[14] DataCamp, https://www.datacamp.com/community/blog/datacamp-donates

[15] ASP2021 workshop for high school teachers, https://indico.cern.ch/event/1050810/

[16] The International Association of Physics Students, https://www.iaps.info/